\documentclass[a4, 10 pt, conference]{ieeeconf}  % Comment this line out if you need a4paper

\IEEEoverridecommandlockouts                              % This command is only needed if 
\usepackage{graphicx} % for pdf, bitmapped graphics files
\usepackage{url}

\title{\LARGE \bf
Dialogue System of Team NTT-EASE for DRC2023
}

\author{Yuki Kubo, Tomoya Yamashita, Masanori Yamada% <-this % stops a space
\thanks{NTT Social Informatics Laboratories, Nippon Telegraph and Telephone Corporation, Tokyo, Japan
{\tt\small yk.kubo@ntt.com}
}% <-this % stops a space
}
\begin{document}

\maketitle
\thispagestyle{empty}
\pagestyle{empty}

%%%%%%%%%%%%%%%%%%%%%%%%%%%%%%%%%%%%%%%%%%%%%%%%%%%%%%%%%%%%%%%%%%%%%%%%%%%%%%%%
\begin{abstract}
We developed a dialogue system as a team NTT-EASE in the Dialogue Robot Competition 2023 (DRC2023).
We introduce a dialogue system (EASE-DRCBot) constructed for DRC2023.
EASE-DRCBot incorporates a manually defined dialogue flow.
The conditions for system utterances are based on keyword extraction, example-based method, and sentiment analysis.
For answering a user's question, EASE-DRCBot utilizes GPT-3.5 to generate responses.
We analyze the results of the preliminary round and explain future works.

\end{abstract}

%%%%%%%%%%%%%%%%%%%%%%%%%%%%%%%%%%%%%%%%%%%%%%%%%%%%%%%%%%%%%%%%%%%%%%%%%%%%%%%%
\section{INTRODUCTION}
In this paper, we present EASE-DRCBot, a dialogue system we developed for the Dialogue Robot Competition 2023 (DRC2023) \cite{DRC, DRC2023}.
In DRC2023, the robot introduce a user a tourist route in a dialogue.

% The whole system of the dialogue robot consists of the architectures made by competition organizers \cite{DRC_system} and our own system.
% EASE-DRCBot has some similar points the system \cite{osbot} which a member of us developed in DRC2022.

Inspired from one of dialogue system~\cite{osbot} in DRC2022, we constructed a dialogue system.
To get high rating in DRC2023, we focus on following points: 
\begin{itemize}
    \item  
    A smooth conversation approximately ten minutes. 
    % We structure a dialogue flow, divided into three manually-designed phases.
    \item
    A conversation which does not occur a dialogue breakdown.
    % EASE-DRCBot analyzes user utterances for system utterances.
    % EASE-DRCBot uses GPT-3.5 to generate responses to a user's question.
\end{itemize}

In the following sections, we explain the features of our system and the design of our dialogue flow, and then we analyze EASE-DRCBot's performance in the preliminary round.

\section{Features of EASE-DRCBot}
We developed a dialogue system combining with the competition organizer's architecture \cite{DRC_system} and DialBB \cite{dialbb}.
In the following, we explain the whole architecture of EASE-DRCBot and main features.
% In the following, we explain the main features of EASE-DRCBot.

\subsection{Whole Architectures}
\begin{figure}[tb]
    \centering
    \includegraphics[width=\linewidth]{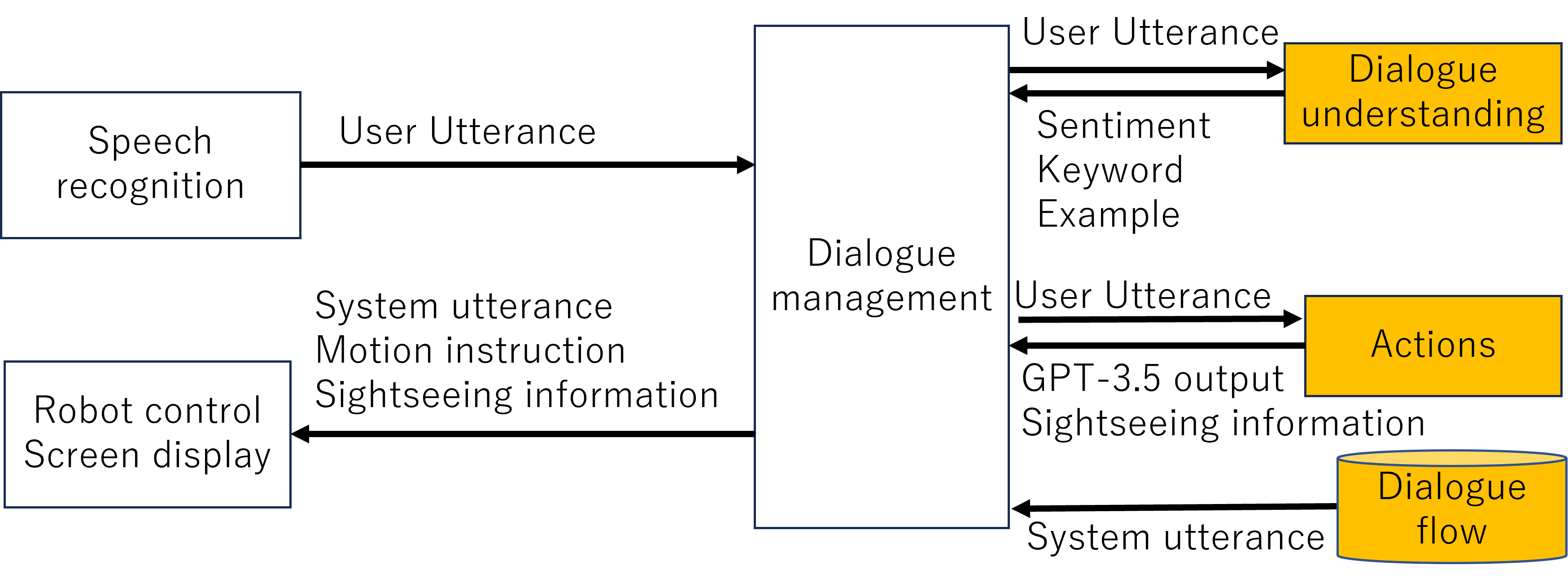}
    \caption{EASE-DRCBot's whole architecture}
    \label{fig:whole_system}
\end{figure}

We show the whole architecture of EASE-DRCBot in Fig.~\ref{fig:whole_system}.
Our developed parts are highlighted orange.
We adopted this architecture for focusing on developing features for system utterances.
% We adopt this architecture to develop system efficiently.

% We developed some functions of prepared architecture in following ways: 
% \begin{itemize}
%     \item EASE-DRCBot interrupts user utterance in six seconds to restrict each turn's time.
%     \item EASE-DRCBot backchannels when utilizing GPT-3.5 to get a user ready to wait our system's response.
% \end{itemize}

The whole processes of our architecture are below:
\begin{enumerate}
    \item A user utterance from the speech recognition, EASE-DRCBot decides a next state from the result of dialogue understanding.
    % \item Our system processes a user utterance into information to use for a state transition condition.
    \item 
    EASE-DRCBot decides a next utterance based on the state, and uses dialogue flow and actions for system utterances. 
    Then, EASE-DRCBot transfers signals to the organizer's architectures.
    % \item Our system decides a next utterance based on information from a user, a described dialogue flow, and described processes.
\end{enumerate}

We describe the methods of outputting a system utterance and the dialogue understanding module.

\subsection{Method of Outputting System Utterance}
We developed the architecture of system utterances combining with DialBB.
We used a dialogue management based on description on spread sheet.
The dialogue management decides a next state and system utterance based on the current state and the result of analyzing a user utterance.
The results of analyzing a user utterance are transferred from the dialogue understanding component.

The following are the conditions for selecting utterances: 
\begin{itemize}
    \item For Yes/No questions, EASE-DRCBot utters based on the result of sentiment analysis and keyword extraction.
    \item For multiple-choice questions, EASE-DRCBot utters based on the result of example-based method and keyword extraction.
    \item In some topics, our system decides to utter a topic by considering previous user utterances.
\end{itemize}

\subsection{Dialogue Understanding}
\begin{table*}[t]
\centering
\caption{
The average result of the scores.
Some question items are same as items in~\cite{DRC}.
}
\label{tab:result}
{\scriptsize
% \begin{tabular}{|l|l|l|l|l|l|l|l|l|l|l|}
% \hline
%                               & Informativeness           & Naturalness               & Appropriateness           & Likeability               & \begin{tabular}[c]{@{}l@{}}Satisfaction\\ with dialogue\end{tabular} & Trustworthiness           & Usefulness                & Correctness               & \begin{tabular}[c]{@{}l@{}}Intention\\ to reuse\end{tabular} & \begin{tabular}[c]{@{}l@{}}Evaluation of\\ travel plan\end{tabular} \\ \hline
% \multicolumn{1}{|r|}{Average} & \multicolumn{1}{r|}{4.30} & \multicolumn{1}{r|}{3.74} & \multicolumn{1}{r|}{4.04} & \multicolumn{1}{r|}{4.43} & \multicolumn{1}{r|}{3.91}                                            & \multicolumn{1}{r|}{4.30} & \multicolumn{1}{r|}{4.48} & \multicolumn{1}{r|}{5.26} & \multicolumn{1}{r|}{4.00}                                    & \multicolumn{1}{r|}{0.65}                                            \\ \hline
% \end{tabular}
\begin{tabular}{l|l|l|l|l|l|l|l|l|l|l|}
\cline{2-11}
                              & Informativeness           & Naturalness               & Appropriateness           & Likeability               & \begin{tabular}[c]{@{}l@{}}Satisfaction\\ with dialogue\end{tabular} & Trustworthiness           & Usefulness                & Correctness               & \begin{tabular}[c]{@{}l@{}}Intention\\ to reuse\end{tabular} & \begin{tabular}[c]{@{}l@{}}Evaluation of\\ travel plan\end{tabular} \\ \hline
\multicolumn{1}{|r|}{Average} & \multicolumn{1}{r|}{4.30} & \multicolumn{1}{r|}{3.74} & \multicolumn{1}{r|}{4.04} & \multicolumn{1}{r|}{4.43} & \multicolumn{1}{r|}{3.91}                                            & \multicolumn{1}{r|}{4.30} & \multicolumn{1}{r|}{4.48} & \multicolumn{1}{r|}{5.26} & \multicolumn{1}{r|}{4.00}                                    & \multicolumn{1}{r|}{0.65}                                           \\ \hline
\end{tabular}
}
\end{table*}
EASE-DRCBot analyzes a user utterance by utilizing a set of specific functions.
Followings are features in dialogue understanding module.
% Following functions are modules of `dialogue understanding'.
% Our system processes user utterances by utilizing a set of specific functions, subsequently generating output results.
% In the `dialogue understanding' module, we have developed following functions: 

\subsubsection*{Keyword extraction}
Our system checks if a specific keyword is in a user utterance in two methods: 
if a defined word is in a user utterance, and if a specific label is in a user utterance.
EASE-DRCBot separating a user utterance into words and labels them by using GiNZA\footnote{\url{https://megagonlabs.github.io/ginza/}}.

% We used two methods in keyword extraction: 
% \begin{itemize}
%     \item Our system checks if some keywords we prepared are contained in a user's utterance.
%     \item  Our system employs GiNZA\footnote{\url{https://megagonlabs.github.io/ginza/}} and check if a word labelled `food' is contained in a user's utterance.
% \end{itemize}

\subsubsection*{Example-based method}
% We implement example-based method considering a keyword is not in a user utterance.
% We utilize Sentence-BERT\footnote{\url{https://huggingface.co/cl-tohoku/bert-base-japanese-whole-word-masking}} and cosine-similarity.
To deal with the case that defined keywords are not included in the user's utterance, we implement example-based method using BERT model\footnote{\url{https://huggingface.co/cl-tohoku/bert-base-japanese-whole-word-masking}} and cosine-similarity.

\subsubsection*{Sentiment analysis}
EASE-DRCBot employ sentiment analysis to determine whether a user utterances is positive or negative.
We utilize BERT model\footnote{\url{https://huggingface.co/koheiduck/bert-japanese-finetuned-sentiment}} for sentiment analysis.
Because some user utterances like ``No thanks'' are not analyzed negative, EASE-DRCBot also uses keyword extraction for some topics.
% For certain topics, we implement keyword extraction to detect specific phrases such as ``No thanks'' or ``All right'' within the sentence.
The keyword extraction is used for dealing with error of sentiment analysis.

\section{Dialogue Flow}
\subsection{Phases of Dialogue Flow}
The dialogue flow is constructed manually to smoothly end the dialogue approximately ten minutes.
The dialogue flow is segmented into three distinct phases.
Below is an overview of the phases.

\subsubsection*{Ice break}
EASE-DRCBot initiates an interaction aimed at reducing a user's nervousness.
It shows a quiz about a sightseeing spot's name, where the user is asked to identify a location displayed on a monitor.
Subsequently, It talks with a user about topics to collect user's information.
These topics includes preferences for food, seasons, festivals, the way of spending holidays, driving, and watching TV.

\subsubsection*{Sightseeing}
% In this phase, our system informs the user that the primary objective of the conversation is to recommend sightseeing spots to engage a user to talk about a travel plan.
EASE-DRCBot informs the user that the primary objective of the conversation is to recommend tourist routes.
It then proceeds to ask some questions to understand the user's travel preferences.
The topics in this phase contains sightseeing spots, how to transfer, food, and festival.
% The topics in this phase contains how many sightseeing to see, how to transfer, whether to visit famous place forecast in TV, what to eat, whether to go to festival.

\subsubsection*{Recommending tourist routes}
% In the Recommending sightseeing phase, our system introduces some sightseeing spots to a user, then answers questions from a user.
EASE-DRCBot shows two tourist routes to encourage a user to choose a better one.
It shows some reasons for recommending the tourist routes using information collected through the conversation.
After showing tourist routes, it answers some questions from a user.
% It uses information contained in the context to show reasons of recommending a tourist route.
% The locations and explanations are described manually.
% We described manually information about the tourist routes.

% By showing the reasons, our system aimed a user to  select the tourist route.
% After showing routes, our system answers some questions from a user until ten minutes have elasped or a user indicates he has no more questions.
% After showing routes, our system answers some questions from a user until the end conditions.
% The conditions are below: 
% \begin{itemize}
%     \item Ten minutes have elapsed.
%     \item A user indicates he has no more questions.
% \end{itemize}

\subsection{System Utterances}
We adopted some tips for reducing dialogue breakdowns and a user's waiting time.
\subsubsection{Inducing user's utterance}
To achieve a smooth conversation, EASE-DRCBot employs strategies to induce user utterances. 
Following is the methods of inducing user's utterance: 
\begin{itemize}
    \item 
    % In its initial utterance, the system requests users to speak clearly to minimize Automatic Speech Recognition (ASR) errors.
    EASE-DRCBot requests users to speak clearly to minimize speech recognition's errors.
    \item 
    At each turn, EASE-DRCBot asks a question for inducing a user to answer that question.
    % At each conversational turn, our system asks a question to a user. This is designed to encourage users to directly answer the question, reducing unexpected responses such as sudden, unrelated questions.
\end{itemize}
\subsubsection{A response using GPT-3.5}
EASE-DRCBot utilizes OpenAI's GPT-3.5 API to answer questions from a user.
EASE-DRCBot says ``Just a minute'' when activating GPT-3.5 for signaling to the user that a response is being prepared.
% We adopted GPT-3.5 on the following considerations: 
% \begin{itemize}
%     \item GPT-3.5 offers quicker response times than GPT-4.
%     \item The complexity of user questions is assumed not high, making GPT-3.5 suitable for our needs.
% \end{itemize}

\section{Results of Preliminary Round}
In TABLE~\ref{tab:result}, we show the average scores for each question.
Each item except ``Evaluation of travel plan'' was measured on a 7-point Likert scale.
Some question items in TABLE~\ref{tab:result} are same as the questions of DRC2022~\cite{DRC}, and some questions mean below: 
\begin{itemize}
    \item ``Correctness'' means ``Do you think the information provided by the robot was correct?''
    \item ``Evaluation of travel plan'' means the ratio of people answered ``Yes'' to both ``Was it able to make a plan to visit two sightseeing spots?'' and ``With one's own common sense, do you think the plan you created is feasible?''
\end{itemize}

From the result, we can say following things: 
\begin{itemize}
    \item
    The scores on the satisfaction on the dialogue (``Naturalness'' and ``Satisfaction with dialogue'') were relatively low. 
    This suggests that we can develop more natural dialogue flow, and that we can improve dialogue understanding module's performance.
    \item 
    The scores on the information uttered by our system (``Usefulness'' and ``Correctness'') were relatively high.
    This suggests that our strategy of presenting two sightseeing spots and explaining the reasons based on the user's information are effective.
\end{itemize}

\section{Conclusion}
We introduced our system EASE-DRCBot which we developed for DRC2023.
EASE-DRCBot utters based on the manually described dialogue flow.
We adopted keyword extraction, example-based method, and sentiment analysis to understand the user's utterance.
By segmenting the dialogue into three distinct phases, we ensured stable dialogue processing. 
EASE-DRCBot used GPT-3.5 to answer the user's question.
From the result, we found that the scores about the dialogue could be improved, and that a more natural dialogue flow is needed for more user satisfaction with the dialogue.

{\scriptsize
\bibliographystyle{IEEEtran}
\bibliography{root}
}
\end{document}